\documentclass{article}
\usepackage{spconf,amsmath,graphicx,bbm}
\usepackage{multirow, hhline, stmaryrd}

\DeclareMathOperator*{\argmin}{arg\,min}

\title{Adversarial Speaker Verification}
\name{Zhong Meng, Yong Zhao, Jinyu Li, Yifan Gong}

\address{Microsoft Corporation, Redmond, WA, USA \\ \normalsize \{zhme, yonzhao, jinyl, ygong\}@microsoft.com}

\begin{document}
\ninept
\maketitle
\begin{abstract}

	The use of deep networks to extract embeddings for speaker recognition has proven successfully. However, such embeddings are susceptible to 
    performance degradation due to the mismatches among the training, enrollment, and test conditions. In this work, we propose an
	adversarial speaker verification (ASV) scheme to learn the
	\emph{condition-invariant} deep embedding via adversarial
	multi-task training. In ASV, a speaker classification network and a condition identification network are jointly optimized to minimize the speaker
	classification loss and simultaneously mini-maximize the condition loss. The target labels of the condition network can be categorical (environment types) and continuous (SNR values).  We further propose multi-factorial ASV to simultaneously
	suppress multiple factors that constitute the condition
	variability.  Evaluated on a Microsoft Cortana text-dependent speaker verification task, the ASV
	achieves 8.8\% and 14.5\% relative improvements in equal error rates (EER)
	for known and unknown conditions, respectively. 

\end{abstract}

\section{Introduction}
Speaker verification (SV) is the task of authenticating the claimed
identity of an utterance, based on a speaker's known recordings. An SV system is text-dependent (TD) if the content of the test
utterance is a fixed or prompted text phrase and is text-independent if the
test utterance is unconstrained speech. Over the years, the i-vector
followed by probabilistic linear discriminant analysis (PLDA) 
\cite{dehak2011front} has been the dominant approach for SV. 

Recently, with the advent of deep learning, deep embeddings learned from a
deep network have achieved great success and have become the
state-of-the-art in speaker recognition. In \cite{variani2014deep, richardson2015deep}, a 
deep neural network (DNN) is trained to discriminate between speakers and the 
outputs from a hidden layer are averaged over frames in an utterances as the  embeddings to represent the enrolled speakers and test utterances. In \cite{snyder2016deep}, pooling over hidden
layer activations in the training time is introduced for speaker
classification. In \cite{heigold2016end,
snyder2016deep}, long short-term memory (LSTM) model is optimized using an end-to-end criterion. A triplet loss is further introduced in
\cite{li2017deep} to learn speaker embeddings.  In \cite{zhang2016end}, the attention mechanism was introduced to determine the weights for
combining the frame-level features, instead of just equally averaging all the frames. Although these methods have greatly advanced the performance of SV, the performance degradation due to mismatched conditions is still a significant barrier for deploying speaker recognition technologies.
The major adverse conditions causing mismatches are different channel and noise environments, and the mismatches exist in two folds, not only between the training and test conditions but also the enrollment and test conditions \cite{snyder2018x, Li14overview}. 


Recently, adversarial learning \cite{gan} has achieved great success in estimating generative models. In speech
area, it has been applied to speech enhancement \cite{pascual2017segan,
donahue2017exploring, meng2018adversarialfeature}, voice conversion
\cite{kaneko2017parallel}, acoustic model adaptation
\cite{meng2019asa, sun2017unsupervised, meng2017unsupervised}, noise-robust
\cite{grl_shinohara, grl_serdyuk}, speaker-invariant \cite{saon2017english,
meng2018speaker} automatic speech recognition, speaker model adaptation
\cite{wang2018unsupervised} and speech enhancement \cite{meng2018cycle, pascual2017segan, meng2018afm} using gradient reversal layer (GRL)
\cite{ganin2015unsupervised}. In these works, adversarial learning is used
to learn an intermediate representation in a DNN that is
invariant to the shift among different conditions (e.g., environments,
speakers, SNRs, etc.). To benefit from this, in this work, we propose
adversarial speaker verification (ASV) to suppress the effects of condition
variability in speaker modeling for robust SV. In ASV, a speaker classification network and a condition identification network are jointly trained to minimize the speaker classification loss and to
mini-maximize the condition loss through adversarial multi-task learning. The target labels of the condition network can be categorical (environment types) and continuous (SNR values)
With ASV, speaker-discriminative and condition-invariant deep
embeddings can be extracted for both enrollment and test speech.

Adversarial learning has been applied to speaker modeling in \cite{wang2018unsupervised}. The proposed ASV is significantly different from it in that:
(1) ASV suppresses two kinds of condition variability in speaker modeling using different methods,
whereas \cite{wang2018unsupervised} aims at adapting a well trained speaker model to the unlabeled target domain data, other than condition robustness of the model. (2) The proposed system train a network
directly takes acoustic features as the input while the input of the system in \cite{wang2018unsupervised} are utterance-level i-vectors which requires additional computational time and resources. 


We perform ASV experiments to normalize environment and SNR
variabilities on a Microsoft Cortana TD-SV task. ASV 8.8\% and 14.5\% relative improvements over the
baseline for known and unknown conditions, respectively. The twice as much relative gain on
\emph{unknown} conditions as on known conditions shows the significant
advantage of ASV in normalizing out the unexpected condition factors in
speaker modeling and the strong capability of generalizing to unknown
conditions.

\section{Deep Embedding for Speaker Verification}
\label{sec:deep_embed}
Deep embedding has been widely used for speaker verification. In the
training stage, we train a background DNN to disriminate between speakers of the training set. 
We view the hidden layers of the background DNN as a feature extractor network
$M_f$ with parameters $\theta_f$ that maps input speech frames
$\mathbf{X}=\{\mathbf{x}_1, \ldots, \mathbf{x}_{T}\}, \mathbf{x}_t\in
\mathbbm{R}^{r_x}, t=1,\ldots, T$ from
training set to intermediate deep hidden features
$\mathbf{F}=\{\mathbf{f}_1, \ldots, \mathbf{f}_{T}\}, \mathbf{f}_t\in
\mathbbm{R}^{r_f}$ and the upper layers of the background DNN as a
speaker classifier network $M_y$ with parameters $\theta_y$ that maps the deep
features $\mathbf{F}$ to the speaker posteriors $p(a|\mathbf{f}_t;
\mathbf{\theta}_y), a\in \mathbbm{A}$
as follows:
\begin{align}
	\mathbf{f}_t & = M_f(\mathbf{x}_t) \\
	p(a | \mathbf{f}_t; \mathbf{\theta}_y) & = M_y(\mathbf{f}_t),
	\label{eqn:speaker_classifier}
\end{align}
where $\mathbbm{A}$ is the set of all speakers in the training set.
$\mathbf{\theta}_f$ and $\mathbf{\theta}_y$ are optimized by minimizing the cross-entropy loss of
speaker classification as below
\begin{align}
	\mathcal{L}_{\text{speaker}}(\theta_f, \theta_y) & = -\frac{1}{T} \sum_{t=1}^{T} \log
	p(y_t | \mathbf{x}_t; \mathbf{\theta}_f, \mathbf{\theta}_y)\nonumber \\
	& = -\frac{1}{T} \sum_{t=1}^{T} \sum_{a\in
		\mathbbm{A}} \mathbbm{1}[a =
		y_t] \log M_y(M_f(\mathbf{x}_t)), \label{eqn:loss_speaker}
\end{align}
where $\mathbf{Y}= \{y_1, \ldots, y_{T}\}, y_t\in \mathbbm{A}$ is a sequence of speaker labels aligned
with $\mathbf{X}$ and $\mathbbm{1}[\cdot]$ is the indicator function which equals to 1 if the condition in the squared bracket is satisfied and 0 otherwise. 
Once the speaker classifier is trained, we compute deep hidden features for given
utterances and average them to form a compact deep embedding for that speaker. Note
that the speakers in the enrollment and training sets do not overlap. During
evaluation, the cosine distance between the embeddings of the test utterance and the
claimed speaker is computed and compared with a threshold to make the
verification decision.

\section{Adversarial Speaker Verification}
\label{sec:asv}
With the background DNN for speaker classification, we are able to learn
speaker-discriminative embeddings.  However, in some scenarios, the speakers are enrolled
in different conditions (i.e., environments, SNR values, etc.) from those
in the training set and the test utterances are recorded in different
conditions from the training and enrollment sets. Under these scenarios,
the embeddings of the enrolled speakers and the test utterances are mismatched and may lead to degraded speaker verification performance because
the new conditions for enrollment and testing are \emph{unknown} to the
background DNN trained on the training set.  We propose ASV to reduce the
effects of condition variability on the background DNN, i.e., the high
variances of hidden and output unit distributions of the network caused by
the inherent inter-condition variability in the speech signal. With such a
background DNN, we can extract \emph{condition-invariant} deep embeddings
for enrolled speakers and test utterances to perform robust SV. 

\subsection{Adversarial Training of Background DNN}
With ASV, our goal is to learn a \emph{condition-invariant} and
\emph{speaker-discriminative} deep hidden feature in the background DNN
through adversarial multi-task learning such that a noise-robust deep
embedding can be obtained from these deep features for a enrolled speaker
or a test utterance. In real application, conditions that affect the
speaker modeling can be represented by either a categorical or a continuous
variable. For example, different kinds of environments can be characterized by a categorical variable while the noise levels, i.e., the SNR values, of the input speech frames in the same or different environments are represented by a continuous variable. We propose different methods to deal with these two types of condition
variability as follows.

 \begin{figure*}[!t]
 	\centering
 	\includegraphics[width=16cm]{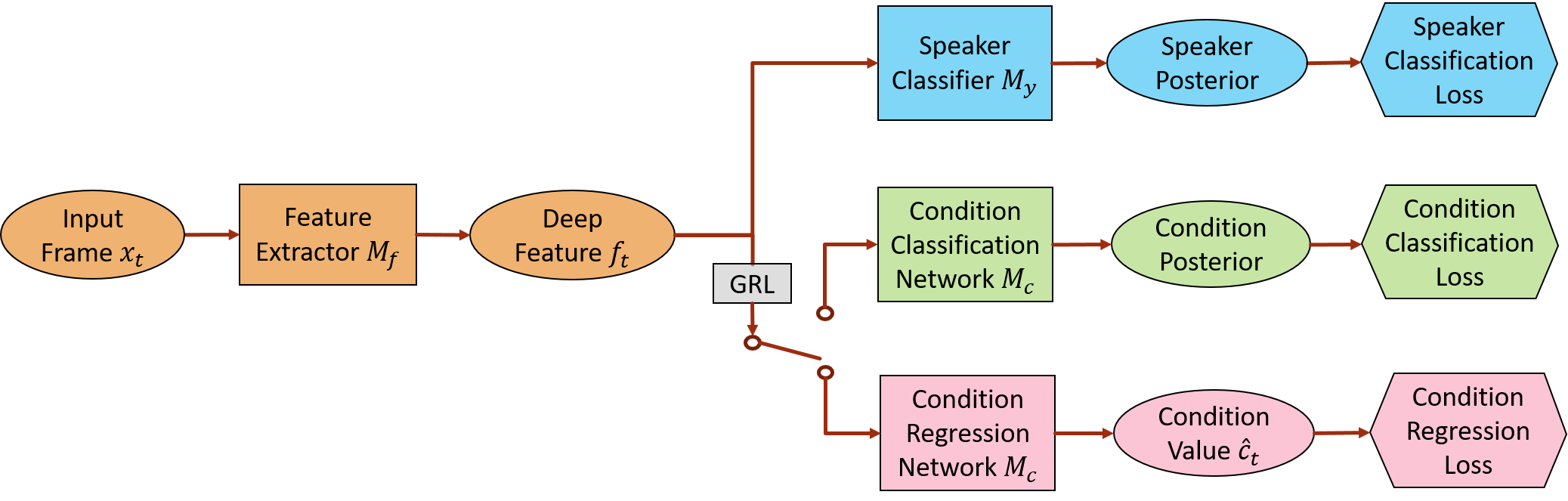}
 	\caption{Adversarial training of background DNN (consisting of $M_f$ followed by $M_y$) for condition-robust speaker verification. The condition \emph{classification} network is selected as the auxiliary network if the conditions can be represented as a categorical variable and the condition \emph{regression} network is used when the conditions are expressed as a continuous variable.
 	}
 	\label{fig:asv}
 \end{figure*}

\subsubsection{Categorical Condition Classification Loss}
As shown in Fig. \ref{fig:asv}, to address the conditions that are 
characterized as a categorical variable, we introduce an additional condition classification network $M_c$
which predicts the condition posteriors $p(b | \mathbf{f}_t; \theta_f), b
\in \mathbbm{B}$ given the deep features $\mathbf{F}$ from the training set as
follows:
\begin{align}
M_c(\mathbf{f}_t) & = p(b | \mathbf{f}_t; \mathbf{\theta}_c) = p(b |
\mathbf{x}_t; \mathbf{\theta}_f, \mathbf{\theta}_c),
	\label{eqn:cond_classify}
\end{align}
where $\mathbbm{B}$ is the set of all conditions  in the training set.
With a sequence of condition labels $\mathbf{C} = \{c_1,
\ldots, c_{T}\}$ that is aligned with $\mathbf{X}$, we are able to compute
the condition classification loss through cross-entropy as follows
\begin{align}
	\mathcal{L}_{\text{condition}}(\theta_f, \theta_c) & = -\frac{1}{T} \sum_{t=1}^{T} \log
	p(c_t | \mathbf{f}_t; \mathbf{\theta}_c)\nonumber \\
	& = -\frac{1}{T} \sum_{t=1}^{T} \sum_{b\in
		\mathbbm{B}} \mathbbm{1}[b =
		c_t] \log M_c(M_f(\mathbf{x}_t)).
	\label{eqn:loss_classify}
\end{align}

\subsubsection{Continuous Condition Regression Loss}
The speaker modeling is also affected by conditions that are
represented by a continuous variable. 
Different from a categorical variable such as the environment type,
continuous conditions are real numbers or real vectors that can hardly be
evaluated by a classification network of which the outputs represent the
posteriors of discrete classes. Therefore, as shown in Fig.
\ref{fig:asv}, we introduce an additional condition regression network
$M_c$ instead to predict the frame-level condition value (e.g., SNR value)
$\hat{\mathbf{c}}_t\in \mathbbm{R}^{r_c}$ given the deep features $\mathbf{F}$
from the training set as follows:
\begin{align}
	M_c(\mathbf{f}_t) & = \hat{\mathbf{c}}_t.
	\label{eqn:snr_regress}
\end{align}
With a sequence of ground truth condition values $\mathbf{C} =
\{\mathbf{c}_1,
\ldots, \mathbf{c}_{T}\}$ that is aligned with $\mathbf{X}$, we are able to compute
the condition regression loss through mean-square error as follows
\begin{align}
	\mathcal{L}_{\text{condition}}(\theta_f, \theta_c) & = -\frac{1}{T} \sum_{t=1}^{T}
	(\hat{\mathbf{c}}_t - \mathbf{c}_t)^{2} \nonumber \\
	& = -\frac{1}{T} \sum_{t=1}^{T} \left[M_c(M_f(\mathbf{x}_t)) - \mathbf{c}_t
	\right]^{2}. \label{eqn:loss_regress}
\end{align}

\subsubsection{Optimization}
\label{sec:optimize}
For brevity, we name both condition classification and regression loss as condition loss
$\mathcal{L}_{\text{condition}}$ and we call both
condition classification and regression networks as condition network
$M_c$.  To make the deep features $\mathbf{F}$ condition-invariant, the
distributions of the deep features from different conditions should be as close
to each other as possible. Therefore, the $M_f$ and $M_c$ are jointly
trained with an adversarial objective, in which $\mathbf{\theta}_f$ is
adjusted to \emph{maximize} the frame-level condition loss
$\mathcal{L}_{\text{condition}}$
while $\mathbf{\theta}_c$ is adjusted to \emph{minimize}
$\mathcal{L}_{\text{condition}}$. This minimax competition will first increase the discriminativity of 
$M_c$ and the speaker-invariance of the deep features generated by $M_f$, and
will eventually converge to the point where $M_f$ generates extremely
confusing deep features that $M_c$ is unable to distinguish.

At the same time, we want to make the deep features speaker-discriminative by
minimizing the speaker classification loss $\mathcal{L}_{\text{speaker}}$ as in Eq. \eqref{eqn:loss_speaker}. 

Overall, we find the optimal parameters $\hat{\theta}_y, \hat{\theta}_f$, $\hat{\theta}_c$ through adversarial multi-task learning as follows  
\begin{align}
    (\hat{\theta}_f, \hat{\theta}_y) &= \argmin_{\theta_f, \theta_y} \mathcal{L}_{\text{speaker}}(\theta_f, \theta_y) - 
	\lambda\mathcal{L}_{\text{condition}}(\theta_f, \hat{\theta}_c) \\
    (\hat{\theta}_c) & = \argmin_{\theta_c} \mathcal{L}_{\text{condition}}(\hat{\theta}_f, \theta_c), 
\end{align}
where $\lambda$ controls the trade-off between the
speaker classification loss and the condition loss in
Eq.\eqref{eqn:loss_speaker} or Eq.\eqref{eqn:loss_classify} respectively.



The optimization can be implemented through standard stochastic gradient descent by inserting a GRL in between the feature extractor and the condition network \cite{ganin2015unsupervised}. This GRL serves as an identity transform in the forward propagation and multiplies the gradient by $-\lambda$ during the backward propagation.

\subsection{Enrollment and Evaluation}

The optimized feature extractor $M_f$ is used 
to generate \emph{condition-invariant} deep embeddings for enrolled speakers and test utterances. We use $\mathbf{h}^s$, the mean of deep features given speech frames of an enrolled speaker $s$ at the input of $M_f$, as the deep embedding for $s$ and use $\mathbf{h}^u$, the mean of deep features generated by feeding a test utterance $u$ into $M_f$, as the deep embedding for $u$. The cosine distance between each pair of $\mathbf{h}^u$ and $\mathbf{h}^s$ is computed as the score and is compared with a threshold to make the speaker verification decision.

Note that we do not perform probabilistic linear discriminant analysis (PLDA) scoring which is widely adopted in most i-vector and deep embedding systems because: (1) The parameters of PLDA classifiers are learned from the training data and may not generalize well to the test data, especially that with \emph{unknown} noise. Cosine distance is a non-parametric metric and is more robust to mismatched noisy conditions during testing. (2) In this work, the extracted deep embeddings are already low-dimensional (e.g., 200) and PLDA does not further improve the SV performance from our experience. In other words, the dimension reduction has been implicitly performed by the background DNN. 

\section{Experiments}
In this work, we perform ASV by suppressing the environment and SNR variabilities
in speaker modeling and evaluate its performance on Microsoft Cortana TD-SV task.

\vspace{-5pt}
\subsection{Dataset Description}
We collect a set of utterances that start with the voice activation phrase ``Hey Cortana'' from the Windows 10 desktop Cortana service logs. ``Hey Cortana'' segments are cut out using a keyword detector. The duration of each keyword segment is around 65 to 110 frames. We select 6.8M utterances from 8k different speakers, each with 100 to 1000 utterances as the training set. 
We then simulate 20.4M noisy utterances by adding 4 types of real noise (on buses (BUS), in caf́es (CAF), in pedestrian areas (PED), at street junctions (STR)) from CHiME-3 \cite{chime3_barker} dataset to the 6.8M clean Cortana data to form the noisy training data. The noise type is randomly selected and the noise level is randomly scaled before simulating each noisy utterance to make sure the amount of noisy data of each type is almost the same and the utterance-level SNR values of simulated data are within the range of 0dB to 20dB. The final training set consists of both the clean utterances and the simulated noisy utterances.

From the clean Cortana data, we select 6 utterances from each of the 3k speakers as the enrollment data (called ``Enroll A''). We select 60k utterances from 3k target speakers and 3k impostors in Cortana dataset and mix them with CHiME-3 real noise to generate the noisy evaluation set (called ``Test A'') in the same way as training data simulation. Enroll A is always used for enrollment when Test A is used for evaluation. Speakers in the training set and Enroll A (or Test A) do not overlap. Since Test A share the same types of noise with  the training set, it is used to evaluate the ASV performance with known conditions. 

Further, we record 4 close-talk clean utterances from each of the 183 speakers using far-field devices to create a new enrollment set (called ``Enroll B''). We use devices of the same type to collect 5546 real noisy test utterances from 183 target speakers and 183 impostors under various environments (e.g., background music, TV, etc.) and with different recording distances (e.g., at 1m and 5m). We call this test set ``Test B''. Enroll B is always used for enrollment when Test B is used for evaluation. There is no overlap between the speakers in the training set and Enroll B (or Test B). With completely different recording conditions, Test B is used to evaluate the generalization capability of ASV to unknown conditions.



\subsection{Baseline System}
As the baseline system, we first train a feed-forward background DNN for speaker
classification using 6.8M utterances from the training set with cross-entropy criterion and
extract the deep embeddings of enrolled speakers and test utterances for
speaker verification as described in Section \ref{sec:deep_embed}.
Our baseline is similar to the x-vector system \cite{snyder2018x} in that data augmentation is applied by adding different types of noise to improve the robustness of deep embeddings.


The 29-dimensional log Mel filterbank features together with 1st and 2nd
order delta features (totally 87-dimensional) are extracted. Each frame is
spliced together with 25 left and 25 right context frames to form a
4437-dimensional input feature. The spliced features are fed as the input of the
feed-forward DNN after global mean and variance normalization. The DNN has
5 hidden layers with 2048, 1024, 1024, 512, 200 hidden units for the bottom to the top hidden layers, respectively. The non-linear activation function for each hidden layer is
relu. The output layer of the DNN has 8398 output units corresponding to
8398 speakers in the training set with softmax non-linearity. The
200-dimensional deep embeddings for enrolled speakers and test utterances
are computed by taking the average of the last hidden layer outputs.  

As
shown in Tables \ref{table:eer_testa} and \ref{table:eer_testb}, the EERs for the deep embedding are
4.22\% and 13.02\% on Test A and Test B, respectively. In Table \ref{table:eer_testb}, Test B is first categorized into Quiet, TV and Music based on the type of noise/interference and is then re-classified according to the recording distance into 1m, 3m and 5m categories. For example, a test utterance can be recorded under background Music with a distance of 3m from the speaker.

\begin{table}[h]
\centering
\begin{tabular}[c]{c|c|c|c|c}
	\hline
	\hline
	System & \hspace{4pt} DE \hspace{4pt} & EI DE & SI DE & EI+SI DE \\
	\hline
	EER (\%) & 4.22 & 3.95 & 3.98 & \textbf{3.85} \\
	\hline
	\hline
\end{tabular}
\caption{The speaker verification EER (\%) of baseline deep embedding (DE), ASV
	with environment-invariant (EI) DE only, ASV with SNR-invariant
	(SI) DE only and ASV with both EI and SI DE on Test A with
        \emph{known} conditions.}
\label{table:eer_testa}
\end{table}


\begin{table}[h]
\setlength{\tabcolsep}{3.5pt}
\centering
\begin{tabular}[c]{c||c|c|c||c|c|c||c}
	\hline
	\hline
	System & Quiet & TV & Music & 1m & 3m &
	5m & Total \\
	\hline
	DE & 10.52 & 15.53 & 11.71 &
	10.06 & 11.98 & 13.52 & 13.02 \\
	\hline
	EI DE & 9.75 & 14.18 & 10.08 & 9.00 & 10.74 & 13.00 &
	 11.71 \\
	\hline
	SI DE & 9.95 & 14.18 & 10.40 &
	\textbf{8.76} & 10.95 & 12.93 & 11.80 \\
	\hline
	EI+SI DE & \textbf{9.32} & \textbf{14.00} &
	\textbf{9.46} &	8.87 & \textbf{10.09} & \textbf{12.72} &	\textbf{11.13} \\
	\hline
	\hline
\end{tabular}
\caption{The speaker verification EER (\%) of baseline deep embedding (DE), ASV
	with environment-invariant (EI) DE only, ASV with SNR-invariant
	(SI) DE only and ASV with both EI and SI DE on Test B with
\emph{unknown} conditions.}
\label{table:eer_testb}
\end{table}

\vspace{-5pt}
\subsection{Adversarial Speaker Verification}

We further perform adversarial training of the baseline background DNN with 6.8M utterances in the training set to learn condition-invariant deep embeddings for ASV.  The
feature extractor $M_f$ is initialized with the input layer and 5 hidden
layers of the background DNN and the speaker classifier network $M_y$ is
initialized with the output layer.  The deep hidden feature is the
200-dimensional output of the last hidden layer of the background DNN.  We
first address the effect of environment variability which is a factor of
categorical conditions. The environment classification network $M_c$ is a
feedforward DNN with 2 hidden layers and 512 hidden units for each layer.
The output layer of $M_c$ has 5 units predicting the posteriors of 4 noisy and 1 clean 
environments in the training set. As shown in Tables \ref{table:eer_testa} and
\ref{table:eer_testb}, with environment-invariant deep embeddings,
the ASV achieves 3.95\% and 11.71\% EERs, which are 6.4\% and 10.1\% relatively
improved over the baseline deep embedding on Test A and Test B, respectively, when $\lambda=1.5$.

Then we explore the reduction of SNR variability, a factor of continuous
conditions. We introduce an SNR regression network $M_c$ designed as a
feedforward DNN with 2 hidden layers and 512 hidden units for each layer.
The output layer of $M_c$ has 1 unit predicting the SNR value of each input speech frame
in the training set. The frames in the same utterance share the same utterance-averaged SNR. With SNR-invariant deep embeddings,
the ASV achieves 3.98\% and 11.80\% EERs, which are 5.7\% and 9.4\% relatively
improved over the baseline deep embedding on Test A and Test B, respectively, when $\lambda=0.002$. Note that the initial SNR regression loss is at
the order of magnitude of 2, while the initial speaker classification loss is at
-1. $\lambda$ needs to be small enough to match the dynamic ranges of the two
losses. We see that environment-invariant deep embeddings
achieve slightly better ASV EER than SNR-invariant ones.


Multi-factor (MF) adversarial learning was proposed in \cite{meng2018adversarial} to simultaneously suppress multiple factors that cause the condition variability. In this work, we perform MFA speaker verification to learn deep embeddings that are both environment-invariant and SNR-invariant.  The EER further
reduces to 3.85\% and 11.13\% with 8.8\% and 14.5\% relative
improvements over the baseline deep embedding on Test A and Test B,
respectively with $\lambda_1 = 1.5$ for enivironment factor and $\lambda_2 =
0.002$ for SNR factor. MF ASV achieves about 5.3\% additional relative gain over the best
single-factor ASV. In all cases, the SV performance is quite stable with the variation of $\lambda$ values.

From Table \ref{table:eer_testb}, we see that MF ASV achieves smaller relative EER improvements under TV and 5m far-field conditions. The reason is that under TV environment, there exists background speech from another speaker in the TV program as the interference, which makes ASV much harder to verify the identity of the claimed speaker. The reverberance effect exists in addition to the background noise when the recording devices are placed 5m from the speakers. MF ASV only addresses noise type and noise level variabilities and does not explicitly tackle the variability of room impulse response, so its gain is relatively smaller under the far-field condition.

The significantly larger relative EER gains on Test B with \emph{unknown}
conditions for all three ASV systems show that ASV can effectively learn a
canonical condition-invariant speaker model with remarkably increased
generalization ability on unknown conditions.


\subsection{Conclusions}
We propose an adversarial speaker verification method in which a background
DNN for speaker classification is jointly trained with a condition network
to learn speaker-discriminative and condition-invariant deep embeddings for
condition-robust SV. A regression network is used to reconstruct the
continuous condition variable. 

ASV achieves 8.8\% and 14.5\% relative improvements over the deep
embedding baseline on Test A with known conditions and Test B with
unknown conditions. Environment-invariant deep embeddings work better than SNR-invariant ones for ASV.
The joint suppression of multiple factors of
condition variability further improves the ASV performance.
The significantly larger gains of ASV for the
\emph{unknown} conditions shows its strong generalization capability.

\vfill\pagebreak

\bibliographystyle{IEEEbib}
\bibliography{refs}

\end{document}